\documentclass[twocolumn,showpacs,preprintnumbers,superscriptaddress,amsthm,amsmath,amssymb]{revtex4}
\usepackage{graphicx}
\usepackage{dcolumn}
\usepackage{bm}
\usepackage{color}
\usepackage{booktabs,longtable}
\usepackage{multirow}
\usepackage{CJK}
\usepackage{ltxtable}
\usepackage{rotating}
\usepackage[colorlinks,linkcolor=blue,anchorcolor=blue,citecolor=red]{hyperref}

\begin{document}

\title{Study of single-particle resonant states with Green's function method}

\author{C. Chen}
\affiliation{School of Physics and Microelectronics, Zhengzhou University, Zhengzhou 450001, China}

\author{Z. P. Li}
\affiliation{School of Physical Science and Technology, Southwest University, Chongqing 400715, China}

\author{Y.-X. Li}
\affiliation{School of Physics and Microelectronics, Zhengzhou University, Zhengzhou 450001, China}

\author{T.-T. Sun}
\email{ttsunphy@zzu.edu.cn}
\affiliation{School of Physics and Microelectronics, Zhengzhou University, Zhengzhou 450001, China}

\date{\today}

\begin{abstract}
The relativistic mean field theory with the Green's function method is taken to study the single-particle resonant states. Different from our previous work~[\textcolor[rgb]{0.00,0.07,1.00}{Phys.Rev.C \textbf{90},054321(2014)}], the resonant states are identified by searching for the poles of Green's function or the extremes of the density of states. This new approach is very effective for all kinds of resonant states, no matter it is broad or narrow. The dependence on the space size for the resonant energies, widths, and the density distributions in the coordinate space has been checked and it is found very stable. Taking $^{120}$Sn as an example, four new broad resonant states $2g_{7/2}$, $2g_{9/2}$, $2h_{11/2}$ and $1j_{13/2}$ are observed, and also the accuracy for the width of the very narrow resonant state $1h_{9/2}$ is highly improved to be $1\times 10^{-8}$~MeV. Besides, our results are very close to those by the complex momentum representation method and the complex scaling method.
\end{abstract}

\pacs{25.70.Ef, 21.10.Pc, 21.60.Jz}
%25.70.Ef  Resonances
%21.10.Pc  Single-particle levels and strength functions
%21.60.Jz  Nuclear density functional theory and extensions

\maketitle

%\textbf{Keywords:} single-particle resonant states, Green's function method, RMF theory

\section{Introduction}
\label{sec:intr}
The single-particle resonant states in the continuum are playing crucial roles in the formation of halos in exotic nuclei~\cite{PRC2003Sandulescu_68_054323}.
For example, studies by the relativistic continuum Hartree-Bogoliubov theory suggested that giant halos can be formed in the neutron-rich Zr and Ca isotopes if more than two valence neutrons occupy the resonant states with low angular momentums~\cite{PRL1998MengJ_80_460,PRC2002MengJ_65_041302}, and the existence of a possible deformed halo in $^{40,42}$Mg and $^{22}$C is mainly decided by the single-particle states around the Fermi surface including the resonant states in the continuum~\cite{PRC2010Zhou_82_011301,PRC2012Li_85_024312,CPL2012LiLL_29_042101,PLB2018SunXX_785_530}.
As a result, the exploration of resonant states are becoming significantly important and attracting more and more attentions.

During the past years, a series of approaches have been taken or developed in the exploration
of the single-particle resonant states.
Some approaches are based on the conventional scattering theories, such as $R$-matrix theory~\cite{PRL1987Hale_59_763,PR1947Wigner_72_29},
$K$-matrix theory~\cite{PRC1991Humblet_44_2530}, $S$-matrix theory~\cite{Book1972Taylor-ScatteringTheor,PRC2002CaoLG_66_024311}, Jost function approach~\cite{PRL2012BNLu_109_072501,PRC2013BNLv_88_024323}, and the scattering phase shift (SPS) method~\cite{Book1972Taylor-ScatteringTheor,PRC2010LiZP_81_034311,SCP2010Li_53_773}.
Besides, some techniques which are used for bound states have also been extended to study the single-particle resonant states, such as the complex momentum representation (CMR) method~\cite{PRC2006Hagen_73_034321,PRL2016Li_117_062502,PLB2020ShiXX_801_135174},
the complex scaling method (CSM)~\cite{PRC1986Gyarmati_34_95,PRC1988Kruppa_37_383,PRL1997Kruppa_79_2217,PRC2006Arai_74_064311,PR1983Ho_99_1,PRC2010JYGuo_82_034318,PRC2014ZLZhou_89_034307,PRC2012QLiu_86_054312,
PRC2014Shi_90_034319}, the complex-scaled Green's function (CGF) method~\cite{PLB1998Kruppa_431_237,PTP2005Suzuki_113_1273,EPJA2017Shi_53_40},
the real stabilization method (RSM)~\cite{PRC2008ZhangL_77_014312}, and the analytical continuation of the coupling constant (ACCC) method~\cite{PRC2005Guo_72_054319,Book1989Kukulin-TheorResonance,PRC1997Tanaka_56_562,PRC1999Tanaka_59_1391,PRC2000Cattapan_61_067301,CPL2001SCYang_18_196,PRC2004ZhangSS_70_034308,
EPJA2007SSZhang_32_43,PRC2012SSZhang_86_032802,EPJA2012SSZhang_48_40,EPJA2013SSZhang_49_77,PLB2014SSZhang_730_30,PRC2015Xu_92_024324}.

The Green's function (GF) method~\cite{SJNP1987Belyaev_45_783,PRB1992Tamura_45_3271,PRA2004Foulis_70_022706,Book2006Eleftherios-GF} is also a successful candidate for studying resonances. It can treat the continuum exactly.
With this method, the single-particle spectrum covering the bound states and the continuum are treated on the same footing, exact energies and widths can be obtained for resonant states of all kinds, and correct asymptotic behaviors are kept well for the density distributions. Besides, it is very convenient to be combined with nuclear models. As a result, Green's function method has been used extensively in the study of the nuclear structure and excitations. For example, by applying the GF method to the Hartree-Fock-Bogoliubov (HFB) theory in the coordinate representation, halos in both spherical and deformed nuclei are described very well~\cite{PRC2009Oba_80_024301,PRC2011ZhangY_83_054301,PRC2012YZhang_86_054318,PRC2019SunTT_99_054316}.
Besides, the continuum quasiparticle random-phase approximation (QRPA) formulated with Green's function method~\cite{NPA2001Matsuo_696_371} is developed to describe many interesting phenomena, such as the collective excitations~\cite{PTPS2002Matsuo_146_110,PRC2005Matsuo_71_064326,NPA2007Matsuo_788_307,PTP2009Serizawa_121_97,PRC2009Mizuyama_79_024313,PRC2010Matsuo_82_024318,PRC2011Shimoyama_84_044317}, monopole pair vibrational modes and associated two-neutron transfer amplitudes~\cite{PRC2013Shimoyama_88_054308}, and neutron capture reactions~\cite{PRC2015Matsuo_91_034604}.

The covariant density functional theory (CDFT)~\cite{ANP1984SerotBD_16_1,RPP1989Reinhard_52_439,PPNP1996Ring_37_193,JPG2015JMeng_42_093101} has got remarkable achievements in describing many systems and interesting phenomena, such as the stable and exotic nuclei~\cite{PPNP2006MengJ_57_470,PRC2017ZhangW_96_054308,CPC2017Zhang_41_094102,PRC2018ZhangW_97_054302},
hypernuclei~\cite{PRC2011BNLu_84_014328,PRC2014BNLu_89_044307,PRC2016TTSun_94_064319,PRC2018LiuZX_98_024316}, neutron stars~\cite{CPC2018SunTT_42_025101,PRD2019SunTT_99_023004}, pseudospin symmetries~\cite{PhysRep2015HZLiang_570_1,JPG2017Lu_44_125104,PRC2017Sun_96_044312,PRC2019Sun_99_034310},
and $r$-process simulations~\cite{PRC2008SunB_78_025806,PRC2009NiuZM_80_065806,PLB2013NiuZM_723_172}.
Thus, in recent years, we applied the Green's function method to the framework of the covariant density functional theory. In 2014, the relativistic mean field theory formulated with the Green's function method (RMF-GF) is developed, and as the first time, it is successfully applied to study the single-neutron resonant states~\cite{PRC2014TTSun_90_054321}. It is also confirmed effective for the proton and $\Lambda$-hyperon single-particle resonant states~\cite{JPG2016TTSun_43_045107,PRC2017Ren_95_054318}. In 2016, the relativistic continuum Hartree-Bogoliubov theory combining the Green's function method (RCHB-GF) is developed by containing the pairing correlation, which can describe the halos very well~\cite{Sci2016Sun_46_12006}.
Very recently, Green's function method is further extended to study the resonances in deformed nuclei by solving a coupled-channel Dirac equation with quadrupole-deformed Woods-Saxon potential~\cite{PRC2020Sun_101_014321}.

In our previous works~\cite{PRC2014TTSun_90_054321,JPG2016TTSun_43_045107,PRC2017Ren_95_054318},
the single-particle resonances are identified by comparing the density of states (DOS) displayed by nucleons moving in the mean-field potentials and those by free particles.
According to the DOS difference between the nucleons and free particles, the energy and width of resonant state are given by reading the position and the full-width at half-maximum (FWHM) of the resonant peak, respectively. In this way, energies and widths can be obtained easily for narrow resonances with good accuracy. However, the accuracy decreases for the wide resonances due to the irregular shapes of the resonant peaks. In our recent work~\cite{PRC2020Sun_101_014321}, a direct but very effective approach is proposed to study the resonant states by searching for the extremes of Green's function in terms of the fact that the resonant states are poles locating in the fourth quadrant of the complex energy plane. In this work, we applied this new approach with Green's function method to study the single-particle resonances based on the RMF theory.

The paper is organized as follows. In Sec.~\ref{sec:Theory}, the Green's function method is given briefly. In Sec.~\ref{sec:Numer}, numerical details are presented. After the results and discussions in Sec.~\ref{sec:resu}, a brief summary is drawn in Sec.~\ref{sec:Sum}.

\section{THEORETICAL FRAMEWORK}
\label{sec:Theory}
%%%%%%%%%%%%%%%%%%%%%%%%%%%%%%%%%%%%%%%%%%%%%%
In the RMF-GF theory~\cite{PRC2014TTSun_90_054321}, the Green's function is applied in the coordinate space to calculate the densities for nucleons and the single-particle spectrum of the Dirac equation. The Dirac equation for nucleons in the RMF theory~\cite{ANP1984SerotBD_16_1,RPP1989Reinhard_52_439,PPNP1996Ring_37_193} is,
\begin{equation}
[\bm{\alpha}\cdot\bm{p}+V(\bm{r})+\beta(M+S(\bm{r}))]\psi_n(\bm{r})=\varepsilon_n\psi_n(\bm{r}),
\label{EQ:Dirac}
\end{equation}
with the nucleon mass $M$, the Dirac matrices $\bm{\alpha}$ and $\beta$, and the scalar and vector potentials $S(\bm{r})$ and $V(\bm{r})$, respectively.

A relativistic single-particle Green's function $\mathcal{G}(\bm{r},\bm{r'};\varepsilon)$ satisfying the following definition is needed to be constructed,
\begin{equation}
[\varepsilon-\hat{h}(\bm{r})]\mathcal{G}(\bm{r},\bm{r}';\varepsilon)=\delta(\bm{r}-\bm{r}'),
\label{Eq:GF_define}
\end{equation}
with $\hat{h}(\bm{r})$ being the Hamiltonian of the Dirac equation~(\ref{EQ:Dirac}).
Starting from Eq.~(\ref{Eq:GF_define}) and taking a complete set of eigenstates $\psi_{n}(\bm{r})$ and eigenvalues $\varepsilon_{n}$, the Green's function can be represented as
\begin{equation}
\mathcal{G}(\bm{r},\bm{r}';\varepsilon)=\sum_n\frac{\psi_{n}(\bm{r})\psi_{n}^{\dag}(\bm{r}')}{\varepsilon-\varepsilon_{n}},
\label{EQ:GF}
\end{equation}
which has a form of a $2\times2$ matrix due to the two components of the
Dirac spinor $\psi_{n}(\bm{r})$,
\begin{equation}
\mathcal{G}(\bm{r},\bm{r}';\varepsilon)=
\left(
 \begin{array}{cc}
  \mathcal{G}^{(11)}(\bm{r},\bm{r}';\varepsilon) & \mathcal{G}^{(12)}(\bm{r},\bm{r}';\varepsilon) \\
  \mathcal{G}^{(21)}(\bm{r},\bm{r}';\varepsilon) & \mathcal{G}^{(22)}(\bm{r},\bm{r}';\varepsilon)
 \end{array}
\right).
\end{equation}

It is noted that the eigenvalues $\varepsilon_{n}$ of Dirac equation are poles of the Green's function in Eq.~(\ref{EQ:GF}). As a result, one can obtain the eigenvalues $\varepsilon_{n}$ by searching for the poles of the Green's function. In practice, following Ref.~\cite{Book2006Eleftherios-GF}, it can be done with the help of the density of states (DOS)~$n(\varepsilon)$,
\begin{equation}
n(\varepsilon)=\sum_n\delta(\varepsilon-\varepsilon_{n}),
\label{EQ:dos}
\end{equation}
which displays like discrete $\delta$-function peaks for bound states at the eigenvalues $\varepsilon=\varepsilon_{n}$ but distributes continuously in the continuum with peaks for resonances. DOSs $n(\varepsilon)$ in a wide energy range will be calculated by scanning single-particle energy $\varepsilon$. One notes that for the continuum, energies $\varepsilon$ are complex $\varepsilon=\varepsilon_r+i\varepsilon_i$ and the energies for the resonant states can be written as $\varepsilon_n = E-i\Gamma/2$ with the resonance energy $E$ and width $\Gamma$.

Taking the imaginary part of the Green's function, the DOSs can be calculated by
the integrals in the coordinate ${\bm r}$ space~\cite{PRC2014TTSun_90_054321}.
For the bound states, it is
\begin{eqnarray}
\label{EQ:Sdos}
&&n(\varepsilon)\\
&=&-\frac{1}{\pi}\int d\bm{r}\mathrm{Im}[\mathcal{G}^{(11)}(\bm{r},\bm{r};\varepsilon+i\epsilon)+\mathcal{G}^{(22)}(\bm{r},\bm{r};\varepsilon+i\epsilon)],\nonumber
\end{eqnarray}
where $``i\epsilon"$ is the introduced positive infinitesimal imaginary part to the single-particle energy $\varepsilon$, with which the $\delta$-function shaped DOSs for bound states are simulated by Lorentzian functions with the FWHM of $2\epsilon$.
For the resonant states, one does not need to introduce the infinitesimal imaginary part$``i\epsilon"$ since the single-particle energy $\varepsilon$ is complex. Besides, when scanning the imaginary part of complex energy $\varepsilon_i$, before and after it crossing the resonant states, the DOSs $n(\varepsilon)$ differ by a minus sign. The DOSs for the resonant states can be written as,
\begin{eqnarray}
\label{Eq:DOSres}
&&n(\varepsilon)=\delta(\varepsilon_r-E)\\
&=&\left\{
   \begin{array}{ll}
   {\displaystyle -\frac{1}{\pi}\int d\bm{r}\mathrm{Im}[\mathcal{G}^{(11)}(\bm{r},\bm{r};\varepsilon)+\mathcal{G}^{(22)}(\bm{r},\bm{r};\varepsilon)]} , &\\ \hbox{ ~~~~~~~~~~~~~~~~~~~~~~~~~~~~~~~~~~~~if $\varepsilon_i>-\Gamma/2$;} \\
     {\displaystyle ~~~\frac{1}{\pi}\int d\bm{r}\mathrm{Im}[\mathcal{G}^{(11)}(\bm{r},\bm{r};\varepsilon)+\mathcal{G}^{(22)}(\bm{r},\bm{r};\varepsilon)]}, &\\
 \hbox{~~~~~~~~~~~~~~~~~~~~~~~~~~~~~~~~~~~~~if $\varepsilon_i<-\Gamma/2$.}
   \end{array}
 \right.\nonumber
\end{eqnarray}
In practice, we calculate DOSs for resonances by scanning the whole complex energy range taking the first equation in Eq.~(\ref{Eq:DOSres}) and they will reverse and become negative when $\varepsilon_i$ over the resonant states. According to those changes, the widths of resonant states $\Gamma/2$ can be determined.

In the spherical case, the Green's function can be expanded as
\begin{equation}
\mathcal{G}({\bm r},{\bm r'};\varepsilon)=\sum_{\kappa m}Y_{\kappa m}(\theta,\phi)\frac{\mathcal{G}_{\kappa}(r,r';\varepsilon)}{rr'}Y_{\kappa m}^{*}(\theta',\phi'),
\end{equation}
where $Y_{\kappa m}(\theta,\phi)$ is the spin spherical harmonic, $\mathcal{G}_{\kappa}(r,r';\varepsilon)$ denotes the radial Green's function, and the quantum number $\kappa$ labeling different partial waves, which can give the values of the angular momentums $l$ and $j$,
\begin{equation}
\left\{
  \begin{array}{ll}
    l=\kappa,j=\kappa-\frac{1}{2}, & \hbox{if $\kappa>0$;} \\
    l=-\kappa-1,j=-\kappa-\frac{1}{2}, & \hbox{if $\kappa<0$.}
  \end{array}
\right.
\end{equation}
Then the DOS for each partial wave $\kappa$ is
\begin{eqnarray}
n_{\kappa}(\varepsilon)&=&-\frac{2j+1}{\pi}\int dr\mathrm{Im}\left[\mathcal{G}_{\kappa}^{(11)}(r,r;\varepsilon)\right.\nonumber\\
&&\left.+\mathcal{G}_{\kappa}^{(22)}(r,r;\varepsilon)\right].
\label{Eq:DOS_kappa}
\end{eqnarray}
Practically, we do the integrals in Eq.~(\ref{Eq:DOS_kappa}) in a finite box and obtain an approximate DOS $n_{\kappa}^{R}(\varepsilon)$ for a fixed $R_{\rm box}$.

Finally, a Green's function $\mathcal{G}_{\kappa}(r,r';\varepsilon)$ with angular momentum $\kappa$ and complex single-particle energy $\varepsilon$ is constructed as~\cite{PRB1992Tamura_45_3271}
\begin{flalign}
\mathcal{G}_{\kappa}(r,r';\varepsilon)=
&\frac{1}{W_{\kappa}(\varepsilon)}\left[\theta(r-r')\phi^{(2)}_{\kappa}(r,\varepsilon)\phi^{(1)\dag}_{\kappa}(r',\varepsilon)\right. \nonumber\\
&\left.+\theta(r'-r)\phi^{(1)}_{\kappa}(r,\varepsilon)\phi^{(2)\dag}_{\kappa}(r',\varepsilon)\right],
\label{EQ:RGF}
\end{flalign}
where $\theta(r-r')$ is the step function, $\phi^{(1)}_{\kappa}(r,\varepsilon)$ and $\phi^{(2)}_{\kappa}(r,\varepsilon)$ are two linearly independent Dirac spinors
\begin{flalign}
\phi^{(1)}_{\kappa}(r,\varepsilon)=\left( \begin{array}{c}
g^{(1)}_{\kappa}(r,\varepsilon) \\
f^{(1)}_{\kappa}(r,\varepsilon)
\end{array} \right),\nonumber\\
\phi^{(2)}_{\kappa}(r,\varepsilon)=\left( \begin{array}{c}
g^{(2)}_{\kappa}(r,\varepsilon) \\
f^{(2)}_{\kappa}(r,\varepsilon)
\end{array} \right),
\end{flalign}
obtained by the Runge-Kutta integrals in the whole $r$ space from the asymptotic behaviors of the Dirac spinors at $r\rightarrow0$ and $r\rightarrow\infty$, respectively, and $W_{\kappa}(\varepsilon)$ is the $r$-independent Wronskian funciton defined by
\begin{equation}
W_{\kappa}(\varepsilon)=g^{(1)}_{\kappa}(r,\varepsilon)f^{(2)}_{\kappa}(r,\varepsilon)-g^{(2)}_{\kappa}(r,\varepsilon)f^{(1)}_{\kappa}(r,\varepsilon).
\end{equation}

Exact asymptotic behaviors in the origin and in the infinity are taken for the Dirac spinor. In particular, it is regular at $r\rightarrow 0$ and satisfies
\begin{eqnarray}
\phi^{(1)}_{\kappa}(r,\varepsilon)
   &\longrightarrow& r\left(
                                                       \begin{array}{c}
                                                         j_l(k r) \\
                                                         \frac{\kappa}{|\kappa|}\frac{\varepsilon-V-S}{k}j_{\tilde{l}}(kr)\\
                                                       \end{array}
                                                     \right),
                                                  \label{Eq:behavior_r0}
 \end{eqnarray}
where $\tilde{l}=l+(-1)^{j+l-1/2}$ is the angular momentum of the small component of the Dirac spinor, $k=\sqrt{(\varepsilon-V-S)(\varepsilon-V+S+2M)}$ is the single-particle momentum for all states, and the spherical Bessel function of the first kind $j_l(k r)$ satisfies
\begin{equation}
j_l(k r)\longrightarrow \frac{(kr)^l}{(2l+1)!!},~~~\text{ when}~~r\rightarrow 0.
\end{equation}

The Dirac spinor at $r\rightarrow\infty$ behaves exponentially decaying for the bound states while oscillating outgoing for the continuum, which can be written uniformly as,
\begin{eqnarray}
\phi^{(2)}_{\kappa}(r,\varepsilon)
 &\longrightarrow&\left(
                                                    \begin{array}{c}
                                                      rk h^{(1)}_l(k r) \\
                                                      \frac{\kappa}{|\kappa|}\frac{rk^2}{\varepsilon+2M}h^{(1)}_{\tilde{l}}(k r) \\
                                                    \end{array}
                                                  \right),
\label{Eq:behavior_rinf}
\end{eqnarray}
with the single-particle momentum $k=\sqrt{\varepsilon(\varepsilon+2M)}$ and the spherical Hankel function of the first kind $h^{(1)}_l(k r)$.
%%%%%%%%%%%%%%%%%%%%%%%%%%%%%%%%%%%%
\section{Numerical details}
\label{sec:Numer}
In this work, to compare the results with those obtained by the previous GF calculations~\cite{PRC2014TTSun_90_054321} and also those by CMR~\cite{PRL2016Li_117_062502}, CSM~\cite{PRC2010JYGuo_82_034318}, RSM~\cite{PRC2008ZhangL_77_014312}, and ACCC~\cite{PRC2004ZhangSS_70_034308} methods, we take the same nucleus $^{120}$Sn as an example, and investigate the single-particle resonant states for neutrons by taking the GF method based on the RMF theory. The energies, widths, and the density distributions in coordinate space for resonant states are given and compared with other methods. Both PK1~\cite{PRC2004WHLong_69_034319} and NL3~\cite{NL3} parameters are taken in these RMF calculations.

The equations in the RMF-GF theory are solved in the coordinate space, with different space sizes $R_{\rm{box}}$ and a step of $dr=0.1$~fm. In Eq.~(\ref{EQ:Sdos}), the infinitesimal imaginary parameter $\epsilon$ is taken as $1\times 10^{-6}$~MeV when calculating DOSs for bound states. In calculating the DOSs $n_{\kappa}^{R}(\varepsilon)$ by scanning energies $\varepsilon$ in the fourth quadrant of the complex energy plane, the energy steps $d\varepsilon$ is taken as $1\times10^{-4}$~MeV for both the real energy part and the imaginary energy part in searching for most of resonances. As a result, the predicted energies and widths of the resonant states by the GF method own an accuracy of $0.1$~keV. Besides, much higher accuracy can be easily achieved by taking smaller energy steps $d\varepsilon$.

%%%%%%%%%%%%%%%%%%%%%%%%%%%%%%%%%%%%%
\section{Results and discussion}
\label{sec:resu}

\begin{figure}[h!]
\includegraphics[width=0.45\textwidth]{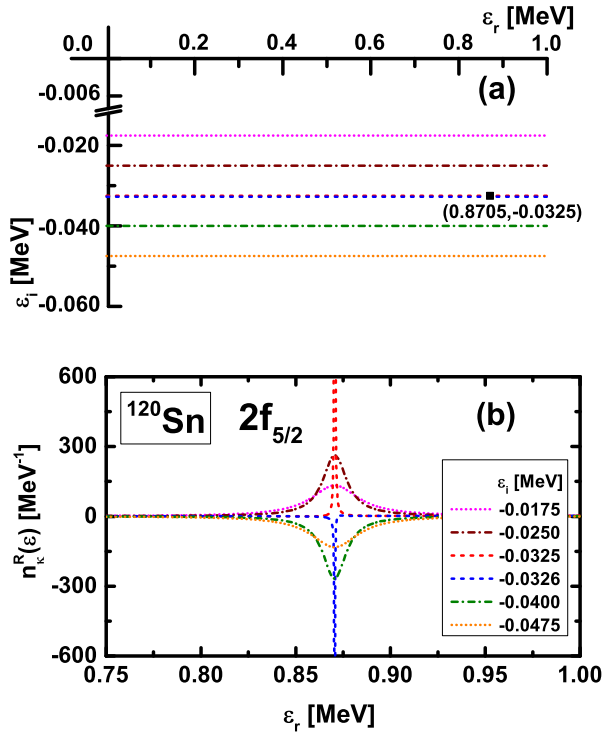}
\caption{(Color online)~(a) Single-particle complex energy plane $\varepsilon=\varepsilon_r+i\varepsilon_i$ and the single-neutron resonant state $2f_{5/2}$ in $^{120}$Sn located in the fourth quadrant. (b) DOSs $n_{\kappa}^{R}(\varepsilon)$ as functions of the complex energy $\varepsilon$ including the real part $\varepsilon_r$ and the imaginary part $\varepsilon_i$, obtained by the RMF-GF method by taking the PK1 effective interaction and space size $R_{\rm box}=20$~fm.}
\label{Fig1}
\end{figure}

The resonant states are well known as poles locating in the fourth quadrant of the single-particle complex energy plane. Therefore, in this work, we take a direct way to explore for these poles which are also the extremes of GF according to Eq.~(\ref{EQ:GF}). In practice, the definition of density of states in Eq.~(\ref{Eq:DOSres}) is applied and a series of DOSs $n_{\kappa}^{R}(\varepsilon)$ will be calculated by scanning the complex energy $\varepsilon$ in the fourth quadrant, both in the directions of the real energy $\varepsilon_{r}$ axis and the imaginary energy $\varepsilon_i$ axis.

\begin{figure}[t!]
\includegraphics[width=0.45\textwidth]{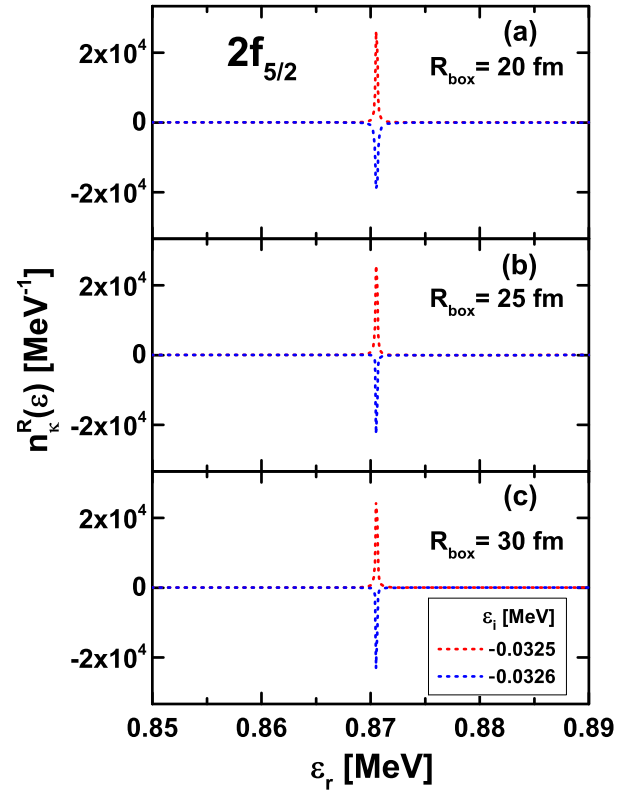}
\caption{(Color online) Comparison of the DOSs $n_{\kappa}^{R}(\varepsilon)$ for the resonant state $2f_{5/2}$ obtained in different space sizes $R_{\mathrm{box}}=20$~fm~(a), 25~fm~(b), and 30~fm~(c), respectively.}
\label{Fig2}
\end{figure}

As an example, in Fig.~\ref{Fig1}, we give the details in determining the single-neutron resonant state $2f_{5/2}$ in $^{120}$Sn. To explore for the pole corresponding to the resonant state, as shown in Fig.~\ref{Fig1}(a), the complex energy $\varepsilon$ in a wide range containing both the real part $\varepsilon_{r}$ and the imaginary part $\varepsilon_{i}$ are covered to calculate the DOSs. The PK1 effective interaction and the coordinate space of $R_{\rm box}=20~$fm are taken \textcolor[rgb]{0.00,0.07,1.00}{in} the RMF-GF calculations. In Fig.~\ref{Fig1}(b), the calculated DOSs are plotted as functions of $\varepsilon_{r}$ for different $\varepsilon_i$. In particular, it is noted that with the imaginary energy $\varepsilon_i$ varying from $-0.0175~$MeV to $-0.0475~$MeV, the DOSs alter significantly in the energy range from $\varepsilon_r=0.75~$MeV to $1.00~$MeV. With the imaginary energy $\varepsilon_i$ approaching to $-0.0325$~MeV, the peaks of DOS evolves sharper and sharper and finally reaches the extreme. A peak in the shape of $\delta$-function locates at $\varepsilon_r=0.8705~$MeV. Besides, just after $\varepsilon_i$ crossing the energy $-0.0325$~MeV, the peak of DOS reverses sharply. After that, the peak of DOS evolves in an apposite way and becomes lower and lower with $\varepsilon_i$ going farther. This indicates a pole locating at $\varepsilon=0.8705-i0.0325~$MeV.

\begin{figure}[t!]
\includegraphics[width=0.45\textwidth]{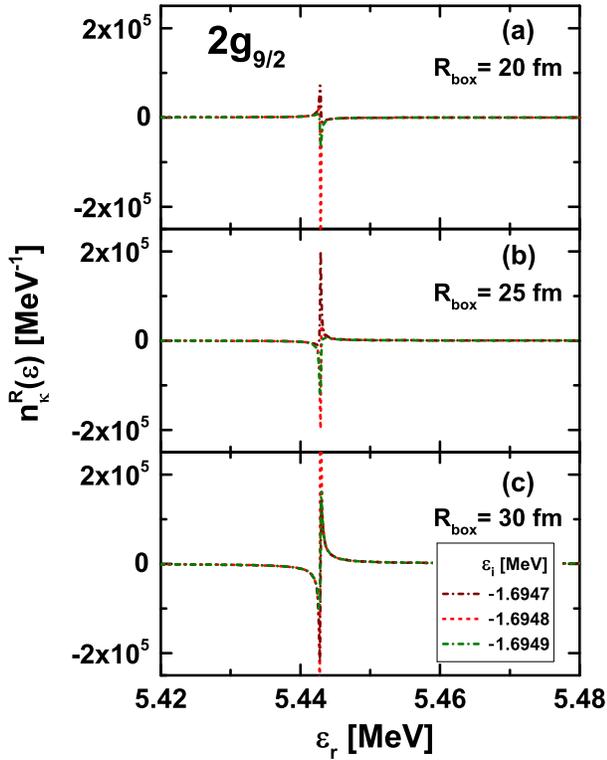}
\caption{(Color online) The same as Fig.~\ref{Fig2}, but for the single-neutron resonant state $2g_{9/2}$ in $^{120}$Sn.}
\label{Fig3}
\end{figure}

In the following, we check the dependence of the obtained resonance energy and width $E-i\Gamma/2$ on the space size,
as we know that they should be constant against the changes of the coordinate space size $R_{\rm box}$. In Fig.~\ref{Fig2}, DOSs calculated by taking different coordinate space sizes $R_{\mathrm{box}}=20$~fm~(a), $25$~fm~(b), and $30$~fm~(c) are plotted for the single-neutron resonant state $2f_{5/2}$ in $^{120}$Sn. It is noted that the shapes of DOSs for $2f_{5/2}$ in different $R_{\rm box}$ are quite similar and all of them reach the extreme at $\varepsilon_i=-0.0325$~MeV and reverse immediately at the next energy point $-0.0326$~MeV. Besides, the peak of DOS in each case also locates at the same energy $\varepsilon_r=0.8705~$MeV. Accordingly, we can conclude that the energy and width of the resonant state $2f_{5/2}$ obtained by the RMF-GF method is independent on the coordinate space size.

The same check plotted in Fig.~\ref{Fig2} is also performed for a wide resonant state. In Fig.~\ref{Fig3}, the DOSs in different $R_{\mathrm{box}}$ are plotted for the resonant state $2g_{9/2}$ with a width around $3$~MeV. In general, for a wide resonant state, the DOS is more sensitive to the changes of imaginary part of complex energy $\varepsilon_i$. In Fig.~\ref{Fig3}, although the DOSs do not have exactly the same shapes with the changes of the space size $R_{\rm box}$, extremes at the same energy $\varepsilon=5.4428-i1.6948$~MeV are observed, demonstrating that the same resonant state with energy $E=5.4428$~MeV and width $\Gamma/2=1.6948$~MeV is obtained in different space sizes. Combing the checks in Figs.~\ref{Fig2} and \ref{Fig3}, it is proved that the descriptions of resonate states by the new approach with the GF method keeps very stable with the changes of the space size, even for a resonant state with broad width.

\begin{figure}[tp!]
\includegraphics[width=0.45\textwidth]{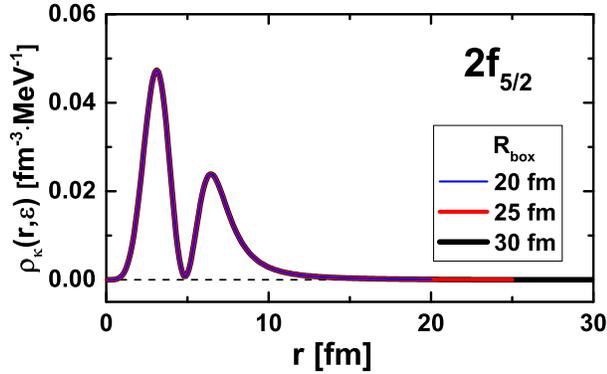}
\caption{(Color online)~Density distributions $\rho_{\kappa}(r,\varepsilon)$ for the single-neutron resonant state $2f_{5/2}$ at resonant energy $\varepsilon= 0.8705$~MeV plotted in the coordinate space. Calculations are done with different space sizes $R_{\mathrm{box}}=20$~fm, 25~fm, and 30~fm.}
\label{Fig4}
\end{figure}

Another advantage of GF method for resonant states is that it can also describe the density distributions in the coordinate space. Here, following Refs.~\cite{PRC2009Oba_80_024301,PRC2011ZhangY_83_054301,PRC2012YZhang_86_054318}, we use the density $\rho_{\kappa}(r,\varepsilon)$ defined at resonance energy $\varepsilon=E$ to describe the distribution for a resonant state in the coordinate space, which is calculated by
\begin{eqnarray}
&&\rho_{\kappa}(r,\varepsilon)\\
&=&-\frac{(2j+1)}{4\pi r^2}\frac{1}{\pi}{\mathrm{Im}}\left[\mathcal{G}_{\kappa}^{(11)}(r,r;E)+\mathcal{G}_{\kappa}^{(22)}(r,r;E)\right].\nonumber
\end{eqnarray}
In Fig.~\ref{Fig4}, the density distribution $\rho_{\kappa}(r,\varepsilon)$ at the resonance energy $\varepsilon=0.8705~$MeV for the state $2f_{5/2}$ in $^{120}$Sn is shown. The space dependence is also checked by doing calculations in different box sizes $R_{\mathrm{box}}=20~$fm, $25~$fm, and $30~$fm. Exactly the same density distribution in the whole coordinate space is obtained with different space sizes, demonstrating again the advantage of GF method. Besides, we can see that the density distribution for the narrow $2f_{5/2}$ is very localized, behaving as a bound state.

\begin{figure}[tp!]
\includegraphics[width=0.45\textwidth]{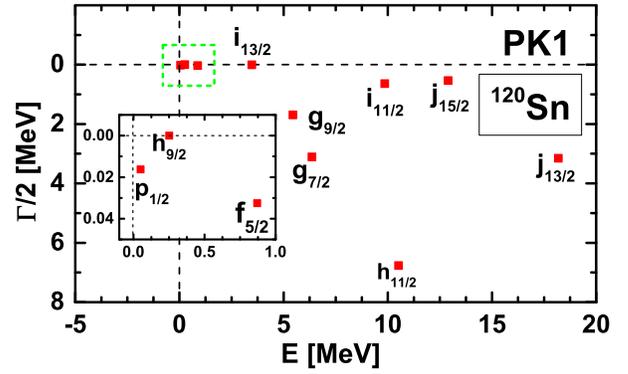}
\caption{(Color online) Single-neutron resonant states in $^{120}$Sn obtained by the RMF-GF method with PK1 effective interaction.}
\label{Fig5}
\end{figure}

\begin{table*}[tp!]
\caption{Energies and widths $E-i\Gamma/2$ (in MeV) of the single-neutron resonant states $nl_j$ in $^{120}$Sn obtained by the GF-RMF method with PK1 effective interaction, compared with the results in the previous GF calculations~\cite{PRC2014TTSun_90_054321}. }
\begin{center}
\begin{tabular}{ccccccc}
\hline \hline
positive parity& $2g_{7/2}$       & $2g_{9/2}$      & $1i_{11/2}$     & $1i_{13/2}$     & &\\\hline
 this work     & $6.3585-i3.1052$ & $5.4428-i1.6948$& $9.8544-i0.6413$& $3.4786-i0.0024$& & \\
 previous work &                  &                 & $9.700-i0.636  $& $3.469-i0.002  $& & \\\hline
negative parity& $3p_{1/2}$       & $2f_{5/2}$      & $1h_{9/2}$      & $2h_{11/2}$    & $1j_{13/2}$ & $1j_{15/2}$ \\\hline
this work      & $0.0504-i0.0164$ & $0.8705-i0.0325$& $0.2508-i4\times 10^{-8}$ & $10.5130-i6.7681$ & $18.1846-i3.1531$ & $12.8929-i0.5322$ \\
 previous work & $0.031-i0.043$   & $0.887-i0.032$  &  $0.251-i0.0001$ &  &  & $12.956-i0.688$ \\\hline \hline
\end{tabular}
\end{center}
\label{Tab1}
\end{table*}

According to the above studies, GF method is effective and reliable in describing resonant states, no matter it is narrow or broad. Resonance energies $E-i\Gamma/2$ can be easily obtained by searching for the poles of Green's function or extremes of DOS. In Fig.~\ref{Fig5}, we plot all the obtained single-neutron resonant states in $^{120}$Sn, identified by scanning the complex energy $\varepsilon$ in a wide range for different $\kappa$
blocks and searching for resonant states by observing extremes. Compared with the results in our previous work (see Fig.6 in Ref.~\cite{PRC2014TTSun_90_054321}), in which the resonant states were identified by comparing the DOSs for nucleons moving in the mean field potentials with those for free particles, new resonant states $2g_{7/2}$, $2g_{9/2}$, $2h_{11/2}$, and $1j_{13/2}$ with very broad widths ranging from $3$~MeV to $13$~MeV are also observed. In Table~\ref{Tab1}, we list the energies $E-i\Gamma/2$ of the single-neutron resonant states obtained by RMF-GF method, in comparison with the results in the previous GF calculations~\cite{PRC2014TTSun_90_054321}. It is found that the accuracy is highly improved with the new approach by GF method, both for the narrow resonant states and broad ones. For example, the uncertainty is well constrained within $1.0\times 10^{-8}$~MeV for the extremely narrow resonant state $h_{9/2}$. It is noted that for the very narrow resonant states, the scanning energy step $d\varepsilon$ for the imaginary part in calculating DOSs $n(\varepsilon)$ should be much smaller. It is $1\times 10^{-8}$~MeV for the resonance $h_{9/2}$, and only with such a small imaginary energy step, the reverse of DOSs extremes can be observed.

\begin{table*}[tp!]
\caption{Energies and widths $E-i\Gamma/2$ (in MeV) of the single-neutron resonant states in $^{120}$Sn obtained by the GF method based on the RMF theory, in comparison with the results by the RMF-CMR, RMF-CSM, RMF-RSM, and RMF-ACCC methods. All calculations are done with NL3 effective interaction.}
\begin{center}
\begin{tabular}{cccccc}
\hline \hline
$nl_{j}$ & GF& CMR~\cite{PRL2016Li_117_062502} & CSM~\cite{PRC2010JYGuo_82_034318} & RSM~\cite{PRC2008ZhangL_77_014312} & ACCC~\cite{PRC2004ZhangSS_70_034308} \\\hline
$2f_{5/2}$   &$ 0.674-i0.015$ &$ 0.678-i0.015$&$ 0.670-i0.010$& $ 0.674-i0.015$  &$ 0.685-i0.012$ \\
$1i_{13/2}$  &$ 3.263-i0.002$ &$ 3.267-i0.002$&$ 3.266-i0.002$& $ 3.266-i0.002$  &$ 3.262-i0.002$ \\
$1i_{11/2}$  &$ 9.601-i0.607$ &$ 9.607-i0.608$&$ 9.597-i0.606$& $ 9.559-i0.602$  &$ 9.600-i0.555$ \\
$1j_{15/2}$  &$12.579-i0.496$ &$12.584-i0.496$&$12.577-i0.496$& $12.564-i0.486$  &$12.600-i0.450$ \\
\hline \hline
\end{tabular}
\end{center}
\label{Tab2}
\end{table*}

Finally, to compare our results with those by CMR~\cite{PRL2016Li_117_062502}, CSM~\cite{PRC2010JYGuo_82_034318}, RSM~\cite{PRC2008ZhangL_77_014312}, and ACCC~\cite{PRC2004ZhangSS_70_034308} methods, we also calculate the resonant states with the RMF-GF method by taking NL3~\cite{NL3} effective interaction. The energies $E-i\Gamma/2$ for the single-neutron resonant states $2f_{5/2}$, $1i_{11/2}$, $1i_{13/2}$, and $1j_{15/2}$ by those methods are listed in Table~\ref{Tab2}.
We can find that the results by the GF method are all consistent with those of other four methods, especially the CMR and CSM methods. In fact, according to our previous study for the resonances in deformed nuclei~\cite{PRC2020Sun_101_014321}, it was found that GF method and CMR can obtain exactly the same energies for most of the resonant states. One possible reason for the slight difference in the present results may come from the mean-field potential obtained in the iteration calculations of RMF theory. Besides, the GF and CMR methods are performed very differently. The GF method is worked in the coordinate space, in which the resonant states are obtained by searching for its poles corresponding to the eigenvalues of the Dirac equation. However, the CMR method is implemented in the momentum space by diagonalizing the Dirac Hamiltonian.

%%%%%%%%%%%%%%%%%%%%%%%%%%%%%%%%%%%%%%%%%%
\section{Summary}
\label{sec:Sum}
%%%%%%%%%%%%%%%%%%%%%%%%%%%%%%%%%%%%%%%%%%
It is well known that the single-particle resonances are playing crucial roles in the structures of exotic nuclei.
Many methods such as CMR, CSM, RSM, and ACCC has been proposed to study resonant states.
In this work, we apply the Green's function method to study the single-particle resonances
based on the RMF theory. Instead of searching for resonant states by comparing the density of states for nucleons in the mean field potentials with those for free particles, a direct and effective approach, that is to search for the extremes of the density of states or the poles of Green's function, is implemented for all kinds of resonant states, either narrow or broad.

Taking $^{120}$Sn as an example, the resonant states are studied with RMF-GF method by taking PK1 effective interaction. The obtained energies and widths are very stable with the change of coordinate space size. The density distributions for resonant states can also be plotted. Compared with our previous work~\cite{PRC2014TTSun_90_054321}, new resonant states $2g_{7/2}$, $2g_{9/2}$, $2h_{11/2}$, and $1j_{13/2}$ with broad widths are identified. Furthermore, the accuracy for the very narrow resonant state $h_{9/2}$ is improved to be $1\times 10^{-8}$MeV. Besides, to compare our results with those by CMR, CSM, RSM and ACCC methods, calculations for $^{120}$Sn by taking NL3 parameter are also done. It is found that results by Green's function method are very close to those by CMR and CSM although they are very different methods.

\begin{acknowledgements}
This work was partly supported by the Physics Research and Development Program of Zhengzhou University (Grant No.~32410217) and the National Natural Science Foundation of China (Grant Nos.~11505157 and 11875225).
\end{acknowledgements}

%\bibliography{bibliography}

\end{document}